\documentclass[12pt,preprint2]{emulateapj}

\shorttitle{Search for Point Sources of Ultra-High Energy Cosmic Rays}
\shortauthors{R.U. Abbasi et al.}

\begin{document}

\title{Search for Point Sources of Ultra-High Energy Cosmic Rays
above $4.0\times 10^{19}$\,eV Using a Maximum Likelihood Ratio Test}

\author{
R.U.~Abbasi,\altaffilmark{1}
T.~Abu-Zayyad,\altaffilmark{1}
J.F.~Amann,\altaffilmark{2}
G.~Archbold,\altaffilmark{1}
R.~Atkins,\altaffilmark{1}
J.A.~Bellido,\altaffilmark{3}
K.~Belov,\altaffilmark{1}
J.W.~Belz,\altaffilmark{4}
S.Y.~BenZvi,\altaffilmark{5}
D.R.~Bergman,\altaffilmark{6}
J.H.~Boyer,\altaffilmark{5}
G.W.~Burt,\altaffilmark{1}
Z.~Cao,\altaffilmark{1}
R.W.~Clay,\altaffilmark{3}
B.M.~Connolly,\altaffilmark{5}
B.R.~Dawson,\altaffilmark{3}
W.~Deng,\altaffilmark{1}
G.R.~Farrar,\altaffilmark{7}
Y.~Fedorova,\altaffilmark{1}
J.~Findlay,\altaffilmark{1}
C.B.~Finley,\altaffilmark{5}
W.F.~Hanlon,\altaffilmark{1}
C.M.~Hoffman,\altaffilmark{2}
M.H.~Holzscheiter,\altaffilmark{2}
G.A.~Hughes,\altaffilmark{6}
P.~H\"{u}ntemeyer,\altaffilmark{1}
C.C.H.~Jui,\altaffilmark{1}
K.~Kim,\altaffilmark{1}
M.A.~Kirn,\altaffilmark{4}
B.C.~Knapp,\altaffilmark{5}
E.C.~Loh,\altaffilmark{1}
M.M.~Maestas,\altaffilmark{1}
N.~Manago,\altaffilmark{8}
E.J.~Mannel,\altaffilmark{5}
L.J.~Marek,\altaffilmark{2}
K.~Martens,\altaffilmark{1}
J.A.J.~Matthews,\altaffilmark{9}
J.N.~Matthews,\altaffilmark{1}
A.~O'Neill,\altaffilmark{5}
C.A.~Painter,\altaffilmark{2}
L.~Perera,\altaffilmark{6}
K.~Reil,\altaffilmark{1}
R.~Riehle,\altaffilmark{1}
M.D.~Roberts,\altaffilmark{9}
M.~Sasaki,\altaffilmark{8}
S.R.~Schnetzer,\altaffilmark{6}
M.~Seman,\altaffilmark{5}
K.M.~Simpson,\altaffilmark{3}
G.~Sinnis,\altaffilmark{2}
J.D.~Smith,\altaffilmark{1}
R.~Snow,\altaffilmark{1}
P.~Sokolsky,\altaffilmark{1}
C.~Song,\altaffilmark{5}
R.W.~Springer,\altaffilmark{1}
B.T.~Stokes,\altaffilmark{1}
J.R.~Thomas,\altaffilmark{1}
S.B.~Thomas,\altaffilmark{1}
G.B.~Thomson,\altaffilmark{6}
D.~Tupa,\altaffilmark{2}
S.~Westerhoff,\altaffilmark{5}
L.R.~Wiencke,\altaffilmark{1}
and A.~Zech\altaffilmark{6} 
}


\altaffiltext{1}{University of Utah,
Department of Physics and High Energy Astrophysics Institute,
Salt Lake City, UT 84112.}

\altaffiltext{2}{Los Alamos National Laboratory, P.O. Box 1663,
Los Alamos, NM 87545.}

\altaffiltext{3}{University of Adelaide, Department of Physics,
Adelaide, SA 5005, Australia.}

\altaffiltext{4}{University of Montana, Department of Physics and Astronomy,
Missoula, MT 59812.}

\altaffiltext{5}{Columbia University, Department of Physics and
Nevis Laboratories, New York, NY 10027: finley@physics.columbia.edu, 
westerhoff@nevis.columbia.edu.}

\altaffiltext{6}{Rutgers --- The State University of New Jersey,
Department of Physics and Astronomy, Piscataway, NJ 08854.}

\altaffiltext{7}{Center for Cosmology and Particle Physics, Department
of Physics, New York University, New York, NY 10003: gf25@nyu.edu.}

\altaffiltext{8}{University of Tokyo,
Institute for Cosmic Ray Research,
Kashiwa City, Chiba 277-8582, Japan.}

\altaffiltext{9}{University of New Mexico,
Department of Physics and Astronomy,
Albuquerque, NM 87131.}

\begin{abstract}
We present the results of a search for cosmic ray point 
sources at energies above $4.0\times 10^{19}$\,eV in the 
combined data sets recorded by the AGASA and HiRes stereo experiments.
The analysis is based on a maximum likelihood ratio test
using the probability density function for each event rather
than requiring an a priori choice of a fixed angular bin size.
No statistically significant clustering of events consistent
with a point source is found.
\end{abstract}

\keywords{cosmic rays --- acceleration of particles ---
large-scale structure of universe}

\section{Introduction}

The world data set of ultra-high energy cosmic rays is currently
dominated by data recorded with the Akeno Giant Air Shower Array
(AGASA) between 1989 and 2003.  AGASA has published arrival directions
of 57 cosmic-ray events above $4.0\times 10^{19}$\,eV obtained up to 
May 2000~\citep{agasa2000}.

The High Resolution Fly's Eye (HiRes) experiment in Utah has operated 
in stereo mode since 1999.  The published AGASA data set and the HiRes 
stereo data set overlap only a few months in time, but both detectors 
observe approximately the same part of the northern sky.  With 27 events 
above $4.0\times 10^{19}$\,eV recorded through January 2004,
the HiRes data set now contributes significantly to the world
ultra-high energy data set:  thus it becomes increasingly interesting 
to search for possible point sources in this {\it combined} data set.  

In combining the two data sets, we need to account for the different 
errors on the individual cosmic-ray arrival directions as well as for the 
different background expectations in the two experiments.  In addition, 
the analysis method should be ``unbinned,'' at least in the sense that 
any error from binning is much smaller than other errors in 
the data.  This means that we do not define a fixed maximum angular
separation for events to form a multiplet.  
Rather, we search over the whole HiRes/AGASA sky for possible point source 
positions. 

For analyzing a data set combining events with very different errors, the 
maximum likelihood method is particularly well suited.  We perform 
a likelihood ratio test of the hypothesis that several events in the skymap 
come from a common source.  In other words, we test whether any given position 
on the sky harbors a source which contributes $n_{s}\geq 1$ 
source events to the data set.  The likelihood of this hypothesis is compared
to the null hypothesis $n_{s}=0$ and this likelihood ratio is maximized using 
$n_{s}$ as a free parameter.  By calculating the likelihood ratio for a 
dense grid of points on the sky, we essentially search the entire sky for the 
most likely position of a source of $n_{s}$ events.  The statistical significance
can be estimated by applying the same method to a large set of random isotropic 
data sets and evaluating what fraction of them have a likelihood ratio which 
is equal to or larger than the ratio observed in the real data described above.

It should be noted that it is not obvious that a study of cosmic ray arrival 
directions will help identify their sources, because charged particles 
suffer deflections in Galactic and extragalactic magnetic
fields. The strengths of these fields are poorly known.  Several authors
have recently published estimates on the size of the expected deflections, 
and the results differ considerably~\citep{sigl2004,dolag2003}.
The point source search reported in this paper could be fruitful only if 
deflections at energies above $4.0\times 10^{19}$ eV are sufficiently weak, 
making cosmic ray astronomy possible.

The AGASA data set has been intensively studied and described in the 
literature~\citep{agasa1996,agasa1999,agasa2000,agasa2003}.  The HiRes 
stereo data set is relatively new, so we will describe it in more detail in 
Section 2.  Section 3 gives a description of the likelihood method applied 
here.  Results from the analysis of the combined data set are described in 
Section 4.  In Section 5, we test the likelihood method with 
simulated data sets, and a discussion follows in Section 6.

\section{The HiRes Detector}

HiRes is a stereo air fluorescence experiment with two sites
(HiRes 1 and 2) at the US Army Dugway Proving Ground in the
Utah desert ($112^{\circ}$\,west longitude, $40^{\circ}$\,north 
latitude, with a vertical atmospheric depth of 
$860\,{\mathrm{g}}/{\mathrm{cm}}^{2}$).
The two sites are separated by a distance of 12.6\,km.

The ultra-high energy cosmic ray flux is small and is a steeply falling power 
law in energy.  Thus experiments at ultra-high energies need a large detector 
volume.  Consequently, the primary cosmic ray particles can not be observed 
directly, since they interact in the upper atmosphere and induce extensive air 
showers with of the order of $10^{10}$ particles for a $10^{19}$\,eV primary.
The properties of the original cosmic ray particle, such as arrival 
direction and energy, have to be inferred from the observed properties of
the extensive air shower.  In HiRes, this is achieved by observing
the fluorescence light produced when particles of the extensive air 
shower interact with nitrogen molecules in the atmosphere.  
This method has the advantage that the shower development in the
atmosphere is imaged and important quantities like the shower size
and the height of the shower maximum can be determined directly.
If the shower is viewed simultaneously by two 
detectors in stereo mode, the arrival direction can be reconstructed 
with an accuracy of less than $1^{\circ}$.  The main shortcoming of the 
technique is the low duty cycle of only about $10\,\%$, as air 
fluorescence detectors can only be operated on dark, moonless nights 
with good atmospheric conditions.

To observe fluorescence light from air showers, the detector at each 
site is made up of several telescope units monitoring different parts 
of the night sky.  With 22 (42) telescopes at the first (second) site, 
the full detector covers about 
$360^{\circ}$ ($336^{\circ}$) in azimuth and $3^{\circ}-16.5^{\circ}$ 
($3^{\circ}-30^{\circ}$) in elevation above horizon.  Each telescope 
consists of a mirror with an area of about $5\,\mathrm{m}^{2}$ area for 
light collection and a cluster of 256 photomultiplier tubes in the focal 
plane.

The shower geometry is determined by a global $\chi^2$ minimization using 
both the timing and pointing information from all tubes.  From measurements 
of laser tracks and stars in the field of view of the cameras we estimate 
that the systematic error in the arrival direction determination is not 
larger than $0.2^{\circ}$, mainly caused by uncertainties in the survey 
of mirror pointing directions.  Various aspects of the HiRes detector and 
the reconstruction procedures are described in~\citet{nim2002,star2002, 
matthews2003}.

For the present analysis, we use stereo data taken between December 1999
and January 2004.  This sample is subject to the same quality cuts 
used in~\citet{apjl2004}.  A minimum track length of 
$3^{\circ}$ in each detector, an estimated angular uncertainty in both 
azimuth and zenith angle of less than $2^{\circ}$, and a zenith angle 
less than $70^{\circ}$ are required.  In addition, the estimated energy 
uncertainty is required to be less than $20\%$ and $\chi^2/{\rm dof}<5$ 
for both the energy and the geometry fit.  A total of 27 events above 
$4.0\times 10^{19}$\,eV pass the selection criteria.

An atmospheric data base built from the reconstruction of laser shots is 
used to correct for the impact of the atmospheric conditions on the 
reconstruction of events.  The data base contains hourly values of parameters 
describing the aerosol content of the atmosphere.  Simulations show that 
after correcting for the aerosol content, the impact on the event geometry 
and energy is small for a wide range of different atmospheric conditions.

The angular resolution of HiRes is determined using simulated showers
generated with CORSIKA 6~\citep{corsika1998} and QGSJET for the first 
interaction.  After the quality cuts described above, 68\,\% of all showers 
generated at $10^{19}$\,eV are reconstructed within $0.57^{\circ}$ 
of the true shower direction.  The angular resolution depends weakly on energy.  
The 68\,\% error radius grows to $0.61^{\circ}$ 
and $0.69^{\circ}$ for showers generated at $4.0\cdot10^{19}$\,eV and $10^{20}$\,eV, 
respectively, because showers at higher energy are farther away, on average.

\section{Analysis}

The maximum likelihood method used here is outlined in~\citet{kinnison1982}; 
a useful and clarifying application to the problem of finding a deficit from 
the position of the moon in a cosmic ray skymap may be found in~\citet{wascko2000}.
For a general description of likelihood methods, see for example~\citet{pdb,meyer1975} 
and references therein. 

The likelihood method can be applied both in searches for emission 
from {\it a priori} selected source locations, and in searches for the most likely
location of such a point source on the whole sky.

Consider first a fixed source location in right ascension and declination: 
$\vec{x}_{s} = (\alpha_s, \delta_s)$.  Given a sample of $N$ cosmic ray events,
we suppose that $n_{s}$ events come from the source location $\vec{x}_{s}$,
and that the remaining $N - n_{s}$ events are random background events.  
If the $i$th event is a source event, then its true arrival direction is 
$\vec{x}_{s}$.  The probability density for finding it at some location 
$\vec{x}$ is given by the function $Q_{i}(\vec{x},\vec{x}_{s})$, derived from 
the angular errors of the event and the angular displacement between 
$\vec{x}_{s}$ and $\vec{x}$.  On the other hand, if the $i$th event is a
background event, then the probability density for finding it at a location 
$\vec{x}$ is given by the function $R_{i}(\vec{x})$, derived from the relative 
exposure of the detector to an isotropic background of cosmic rays. The subscript 
is necessary because $R$ depends on whether event $i$ is a HiRes or AGASA event.
Each of these functions is normalized to unity over positions $\vec{x}$ in the sky.

We do not hypothesize which individual events are source or background
events.  We only suppose that there are $n_s$ events in the sample that 
come from some source position $\vec{x}_{s}$.
Therefore, the partial probability distribution of arrival directions $\vec{x}$
for the $i$th event is given by:
\begin{equation}
P_i(\vec{x},\vec{x}_{s}) = \frac{n_{s}}{N}\,Q_i(\vec{x},\vec{x}_{s})
                + \frac{N-n_{s}}{N}\,R_i(\vec{x})~~.
\end{equation}
It follows that the probability of finding the $i$th event at the location 
$\vec{x}_{i}$ (where it is actually observed) is $P_i(\vec{x}_{i},\vec{x}_{s})$.  

The likelihood for the entire set of $N$ events is then given by:
\begin{equation}
{\mathcal L}(n_{s},\vec{x}_{s}) = \prod_{i=1}^{N} P_{i}(\vec{x}_{i},\vec{x}_{s})~~.
\end{equation}
The best estimate for the number of source events, under the 
assumption of a point source located at $\vec{x_s}$, is 
determined by finding the value of $n_{s}$ which maximizes $\mathcal{L}$.

Because the value of the likelihood function depends on the 
number of events, a more useful quantity than $\mathcal{L}$ 
is the likelihood ratio $\mathcal{R}$:
\begin{eqnarray}
{\mathcal R}(n_{s},\vec{x}_{s}) & = & \frac{{\mathcal L}(n_{s},\vec{x}_{s})}
        {{\mathcal L}(0,\vec{x}_{s})}\nonumber \\
    &  = & \prod_{i=1}^{N}~
        \left\{\frac{n_{s}}{N}
           \left(\frac{Q_{i}(\vec{x}_{i},\vec{x}_{s})}
                      {R_{i}(\vec{x}_{i})}
           -1\right)
        +1\right\}
\end{eqnarray}
where ${\mathcal L}(0,\vec{x}_{s})$ is the likelihood function of the 
{\it null hypothesis} ($n_{s}=0$).  In practice, we maximize $\ln\mathcal R$,
which is equivalent to maximizing $\mathcal{L}$.

The method described so far is sufficient for testing a specific source
position $\vec{x}_{s}$.  To search the entire sky for the source position
with the strongest signal, we
calculate the likelihood ratio $\ln\mathcal R$ for a dense grid of points 
on the sky covering the full range of equatorial coordinates accessible 
to AGASA and HiRes.  The source position is
essentially treated as a free parameter, along with the number
of source events $n_{s}$.  Searching for the parameters
$\alpha_{s}$, $\delta_{s}$, and $n_{s}$ which maximize the likelihood
ratio will therefore give us the best estimate for the 
position of the source and the number of events it contributes.

The search proceeds as follows.  The visible region of the sky
is divided into a fine grid of points with separations 
$0.1^{\circ}/\cos\delta$ and $0.1^{\circ}$ in $\alpha$ and $\delta$, 
respectively.  Each point is treated in turn as a source location
$\vec{x}_{s}$.  At each point, the specific quantities
$Q_{i}(\vec{x}_{i},\vec{x}_{s})$ and $R_{i}(\vec{x}_{i})$ are
required for every event.
The source probability density function $Q_{i}(\vec{x}_{i},\vec{x}_{s})$ 
depends on the angular resolution associated with the $i$th event,
which in principle may depend on several quantities including
energy, zenith angle, etc.
The background probability density function $R_{i}(\vec{x}_{i})$ 
depends on the detector exposure
to different parts of the sky: it is generally the same function for all
events observed by a given detector, but may in principle be a function
of e.g.\@ energy as well.  For each event, 
$Q_{i}(\vec{x}_{i},\vec{x}_{s})$ needs
to be reevaluated at every grid point, while $R_{i}(\vec{x}_{i})$ 
needs only to be evaluated one time.

Once the values of $Q_{i}$ and $R_{i}$ are specified,
the log likelihood ratio $\ln \mathcal R$ is maximized with respect to $n_s$,
where $n_{s}\geq 0$.
This process is repeated for each position $\vec{x}_{s}$ on the grid.
For most locations, $\ln\mathcal R$ is zero, but 
local maxima will occur in the vicinity of one or more events.  
Note that while 
the method is in some sense binned due to the discrete array of grid points, 
the spacing is chosen small enough so that errors 
introduced by binning are insignificant.

\section{Results}

The procedure described above is applied to the combined data set of 57 
AGASA events and 27 HiRes stereo events above $4.0\times 10^{19}$\,eV.  
The analysis is restricted to the AGASA field of view with 
$-10^{\circ}<\delta<80^{\circ}$, and all but one HiRes event fall into 
this declination range.

\begin{figure}
\plotone{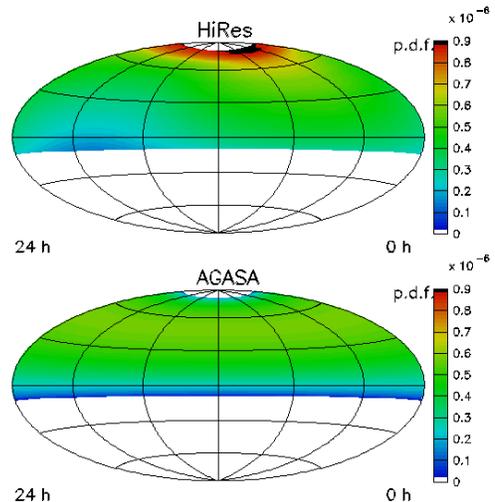}
\caption{Normalized probability densities (p.d.f.) for random background in HiRes and
and AGASA, in equatorial coordinates.  
The sky is binned in $0.1^{\circ}/\cos\delta$ and $0.1^{\circ}$ bins in
right ascension $\alpha$ and declination $\delta$, respectively.
\label{background}}
\end{figure}                                                                                    

The background expectations for HiRes and AGASA are different.
The HiRes exposure has some dependence on right ascension.
Fig.\,\ref{background} shows the normalized HiRes and AGASA 
background map in equatorial coordinates.  

For the signal probability density function (Q) of the HiRes events, we use a 
two-dimensional Gaussian function whose width is chosen such that $68\,\%$ of the 
probability density function falls within an opening angle $0.6^{\circ}$.
Note that for a two-dimensional Gaussian distribution the opening angle 
$\theta=1.515\,\sigma$ encloses $68\,\%$ of the distribution.  Since the 
dependence on energy is weak, we use the same value, $\sigma=0.4^{\circ}$, for
every HiRes stereo event.

For AGASA, we approximate the probability density by the sum of two 
Gaussian functions chosen such 
that the $68\,\%$ and $90\,\%$ opening angle given in~\citet{agasa1999} 
is correctly reproduced,
\begin{eqnarray}
Q & = & \frac{1}{3} \left\{\frac{1}{2\pi\sigma_{1}^{2}} \exp\left(-\frac{(\Delta\theta)^{2}}
     {2\sigma_{1}^{2}}\right)\right\}\nonumber \\
&& +\frac{2}{3} \left\{\frac{1}{2\pi\sigma_{2}^{2}} \exp\left(-\frac{(\Delta\theta)^{2}}
     {2\sigma_{2}^{2}}\right)\right\}~~.
\end{eqnarray}
The width of the Gaussians as a function
of energy has the same energy dependence used in~\citet{stokes2004},
and is given by
\begin{eqnarray}
\sigma_{1} & = & \left(6.52^{\circ} - 2.16^{\circ} 
                   \log(E_{\mathrm{EeV}})\right)/1.515~~~\mathrm{and} \\
\sigma_{2} & = & \left(3.25^{\circ} - 1.22^{\circ} 
                   \log(E_{\mathrm{EeV}})\right)/1.515~~.
\end{eqnarray}
More detailed information on the error shape of individual AGASA events 
could easily be implemented if it becomes available.

\begin{figure*}
\plotone{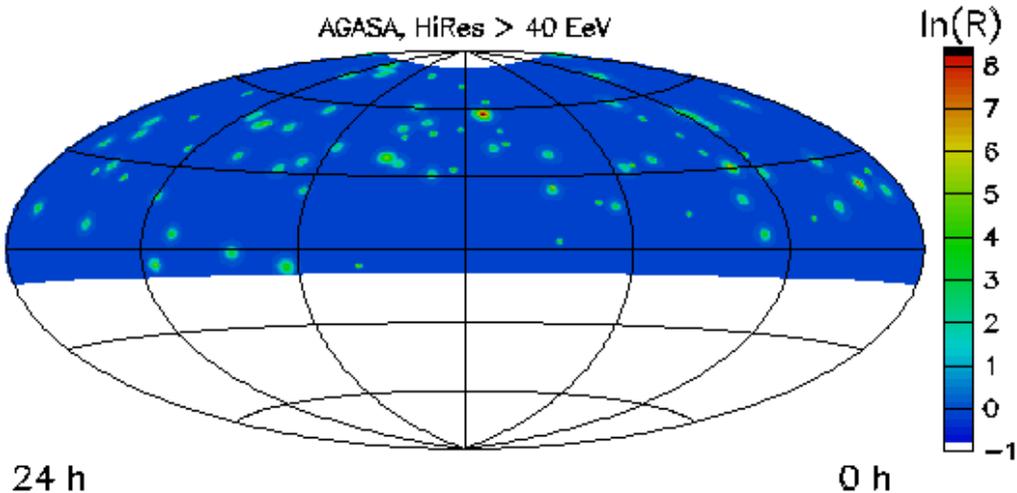}
\caption{Likelihood ratio $\ln\mathcal R$, maximized with respect to $n_{s}$, 
as a function of right ascension and declination of the source position
for the combined set of AGASA and HiRes events above $4.0\times 10^{19}$\,eV.  
Local maxima occur wherever events or clusters of events are located on the
sky.  The global maximum, {\it i.e.} the most likely position of a
``point source'' is at right ascension $\alpha=169.3^{\circ}$ and declination 
$\delta=57.0^{\circ}$.
\label{skymap_40}}
\end{figure*}

Fig.\,\ref{skymap_40} shows the result of the analysis.  At each
$\alpha$ and $\delta$, the likelihood ratio is shown for the
number of source events $n_{s}$ which maximizes $\ln\mathcal R$.
One can clearly recognize where events are located, and one can also 
recognize locations with several nearby events.  AGASA and HiRes 
events can easily be distinguished, as the latter have better 
resolution and therefore smaller regions with large likelihood.  
The point with the largest $\ln\mathcal R$ is at right ascension 
$\alpha=169.3^{\circ}$ and declination $\delta=57.0^{\circ}$.  
The corresponding event cluster comprises 3 nearby AGASA events with
coordinates $(\alpha,\delta)$ and energies $E$ of
(1) $(168.5^{\circ}, 57.6^{\circ})$, $E=77.6$\,EeV, 
(2) $(172.3^{\circ}, 57.1^{\circ})$, $E=55.0$\,EeV, and
(3) $(168.3^{\circ}, 56.0^{\circ})$, $E=53.5$\,EeV. 
This cluster has been described in~\cite{agasa1999} and is listed
as cluster C2 in~\citet{agasa2000}.  The maximum likelihood 
ratio at this position is $\ln\mathcal R=8.54$ 
for $n_{s}=2.9$.  

The behavior of $\ln\mathcal R$ as a function of the fit parameters 
(number of source events and position of the source) in the vicinity 
of the maximum gives error estimates for the fit parameters.  To a good
approximation, the $1\,\sigma$ error is given by the interval over 
which $\ln\mathcal R$ drops by 0.5 from its maximum value.

\begin{figure}
\plotone{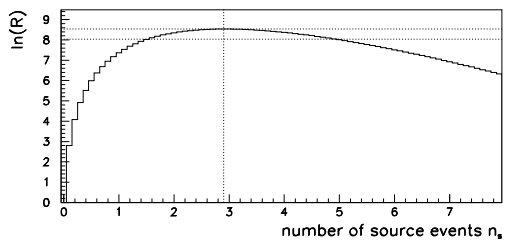}
\caption{Logarithm of the likelihood ratio as a function of the number 
of source events for the position of the maximum.
\label{lnR_vs_n_40}}
\end{figure}                                                                                    

Fig.\,\ref{lnR_vs_n_40} shows $\ln\mathcal R$ as a function
of the number of source events at the position of the maximum.  
The best estimate of the number of source 
events is $n_{s}=2.9^{+2.0}_{-1.4}$.  Similarly we find 
$\alpha=169.3^{\circ}\pm1.0^{\circ}$ and 
$\delta=57.0^{\circ}\pm0.5^{\circ}$ as the best estimate for 
the position of the maximum.

The statistical significance of the appearance of a ``source'' with a 
maximum likelihood ratio $\ln\mathcal R$ in the combined AGASA/HiRes data 
set can be evaluated using simulated random data sets.  The full likelihood 
analysis is performed for $10^{4}$ random data sets with the same number 
of AGASA/HiRes events and the same underlying exposure as the original data 
set, but isotropic arrival directions.  The chance probability for the
``source'' to appear is then given by the fraction of random data sets which 
have at least one location causing the maximum $\ln\mathcal R$ to be equal 
or larger than $8.54$, the value of the maximum in the real data.

\begin{figure}
\plotone{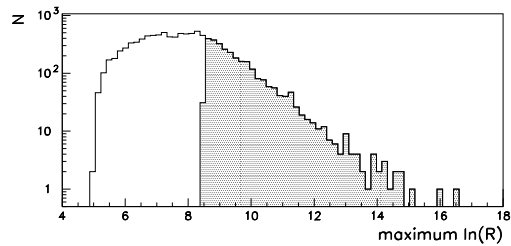} \caption{Maximum $\ln\mathcal R$ for $10^{4}$
simulated random data sets with the same number of AGASA/HiRes events as the
actual data set.  The hatched area marks random sets whose maximum
$\ln\mathcal R$ exceeds the value for the real data set.
\label{lnR_40}}
\end{figure}

Fig.\,\ref{lnR_40} shows the distribution of the maximum $\ln\mathcal R$ for 
each these random data sets.  Out of $10^{4}$ simulated 
data sets, 2793 have a maximum $\ln\mathcal R$ exceeding 
that of the real data set.  The chance probability of the source hypothesis is
therefore of the order of $28\,\%$.  Consequently, there is no
statistically significant evidence for clustering consistent
with a point source in the combined data set.

Note that this is {\it not} simply the chance probability for a 
triplet, but rather the chance probability for a set of 27 HiRes events 
and 57 AGASA events to contain a ``hot spot'' with as high a probability 
to be a ``source'' as the triplet.  Many of the simulated likelihood ratios 
larger than 8.54 in Fig.\,\ref{lnR_40} are indeed caused by doublets.

\section{Test of the Method with Simulated Data Sets}

The maximum likelihood method is tested by applying it to simulated
data sets with sources.  These simulated data sets have $m$ events 
from a common source added to an otherwise isotropic arrival 
direction distribution.  To create such a source, we pick a point
in the sky for the source location and generate $m$ events with 
arrival directions deviating from the source location according to
the probability density function of the individual event as
described in Section 2.  The source location is chosen at random,
but the distribution of locations is forced to reflect the overall
exposure of the detector (Fig.\,\ref{background}), so that
regions with higher exposure are more likely to contain a source.
The $m$ source events replace events in the data set, so the total
number of events is always 83.  There might be additional events
close to the source location due to chance.

\begin{figure}
\plotone{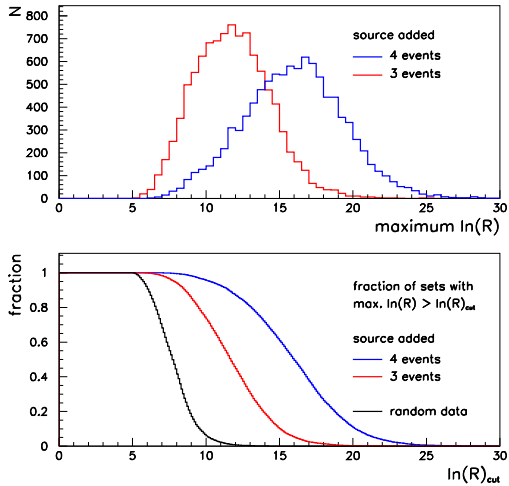}
\caption{{\it Top:} Maximum $\ln\mathcal R$ for $10^{4}$
simulated random data sets where a 4-event source (blue line) or
a 3-event source (red line) has been added.
{\it Bottom:} Fraction of simulated data sets with maximum
$ln\mathcal R > ln\mathcal{R}_{cut}$ as a function of
$ln\mathcal{R}_{cut}$ for simulated random isotropic 
data sets with no source added (black line), a 4-event
source added (blue line), and a 3-event source added
(red line).
\label{eff_40}}
\end{figure}

The full likelihood analysis is applied to these random data sets.
Fig.\,\ref{eff_40}\,(top) shows the maximum $\ln\mathcal R$ for 
$10^{4}$ random data sets where a source with $m=4$ (blue line) or
$m=3$ (red line) has been added.  Fig.\,\ref{eff_40}\,(bottom) shows 
the fraction of simulated data sets with maximum 
$\ln\mathcal R > \ln\mathcal{R}_{cut}$ 
as a function of $\ln\mathcal{R}_{cut}$ for random isotropic data sets with no 
source added ($m=0$, black line), a 4-event source added (blue line), 
and a 3-event source added (red line).
While there is substantial overlap between the distributions, 
the plots also show that point sources which add three or more
events to an isotropic map are recognized with high efficiency.
The medians of the three maximum $\ln\mathcal R$-distributions ($m=0,3,4$)
are 7.7, 11.6 and 16.0, respectively.

\begin{figure}
\plotone{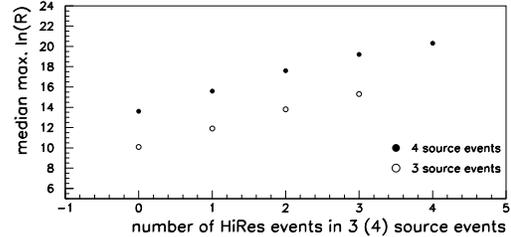}
\caption{Median of the maximum $\ln\mathcal R$-distribution for simulated
data sets with $m=3$ and $m=4$ source events as a function of the 
number of HiRes events contributing to the source cluster.
\label{median}}
\end{figure}                                                                                    

The HiRes and AGASA events in the sample have rather different angular
errors, and we expect this to be reflected in the likelihood ratio
of the simulated data sets.  The maximum $\ln\mathcal R$-distribution for
clusters that are dominated by HiRes events should have a larger
median than the data sets dominated by AGASA events.  Fig.\,\ref{median} 
shows the median of the maximum $\ln\mathcal R$-distribution
as a function of the number of HiRes events that contribute 
to the cluster.  Both for data sets with $m=3$ and $m=4$, the median 
increases if more HiRes events are part of the cluster.

\begin{figure}
\plotone{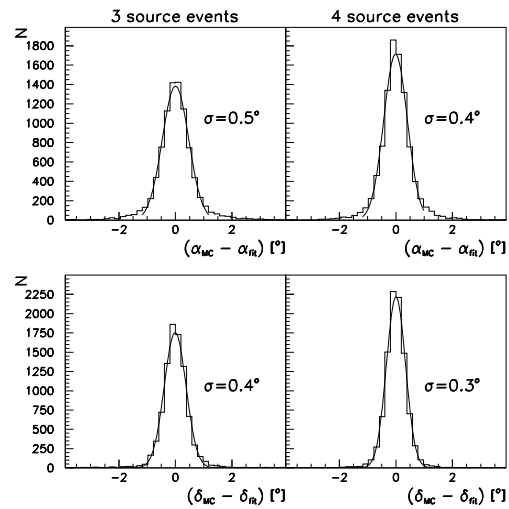}
\caption{Difference between the fitted and true right ascension $\alpha$
and declination $\delta$
for $10^{4}$ data sets with an artificial point source added to an
isotropic data set for $m=3$ (left) and $m=4$ (right) source
events.  The width of a Gaussian fitted to the distribution
is indicated.
\label{reso}}
\end{figure}

Artificial sources can also be used to test the accuracy with which the
position of the point source is reconstructed.  Fig.\,\ref{reso} shows the
difference between the fitted and the true right ascension $\alpha$ and
declination angle $\delta$ for $m=3$ and $m=4$.  
The distributions are fit to a Gaussian function, and the width, averaged
over all source locations and combinations of AGASA and HiRes source events,
is of order $\sigma=0.3^{\circ}$ to $\sigma=0.5^{\circ}$.  The resolution 
improves with the number of events contributing to the signal, as expected.

\section{Discussion}

The chance probability of the triplet using AGASA data alone has
been estimated in~\citet{agasa2003} as being of order $1\,\%$.
This estimate is based on a fixed bin size of $2.5^{\circ}$. 
To test what chance probability an unbinned analysis gives, we
repeat the likelihood analysis for the data set comprising only the
57 AGASA events above $4.0\times 10^{19}$\,eV.  The largest 
likelihood ratio ($\ln\mathcal R=9.66$) appears again near the
events forming the triplet, with
$\alpha=169.3^{\circ}\pm 1.0^{\circ}$ and
$\delta=57.0^{\circ}\pm 0.5^{\circ}$ for $n_{s}=2.9^{+2.0}_{-1.4}$.

As before, we evaluate the chance probability for the appearance
of a source with maximum $\ln\mathcal R=9.66$ or higher in this data set 
by analyzing a large number of simulated isotropic data set, now 
containing 57 AGASA events.  452 out of $10^{4}$ random data sets have
a maximum $\ln\mathcal R$ in excess of 9.66, so the chance probability 
is $4.5\,\%$.  The increase over the chance probability given 
in~\citet{agasa2003} reflects the fact that the unbinned maximum 
likelihood analysis evaluates the quality of the cluster on a continuum.  
This improves the sensitivity to a true point source, and at the
same time it removes the artificial fluctuations that arise in
a binned analysis when an event falls just inside or outside the bin.

The maximum likelihood method allows a straightforward evaluation of the 
chance probability that a given cluster of events come from a common 
source rather than being caused by random background.  It can also be 
used to estimate the chance probability that one or more pre-defined 
positions on the sky correlate with cosmic ray arrival directions.
If several sources contribute, but none of them are strong enough to 
produce a notable value of $\ln\mathcal R$, then the likelihood method 
described here has to be complemented by other methods, such as the angular 
two-point correlation function.  Searches for excess clustering have been 
performed using both the AGASA and HiRes data sets separately, and for 
the combined set.  No significant clustering is reported in the HiRes 
stereo~\citep{apjl2004} and monocular data sets~\citep{stokes2004},
and earlier reports that the AGASA data set shows significant 
clustering~\citep{agasa1999} have recently been questioned.  An unbiased 
analysis~\citep{app2004} which sets aside the fraction of the AGASA data 
that was used to formulate the clustering hypothesis, finds that the 
clustering is consistent at the 8\,\% level with the null hypothesis of 
isotropically distributed arrival directions.  In addition, a two-point 
correlation analysis of the combined AGASA and HiRes data sets finds no 
significant evidence for clustering~\citep{finley2004}.

The energy threshold of $4.0\times 10^{19}$\,eV chosen in this analysis is 
dictated by the fact that only AGASA data above this threshold is published.  
It has been suggested (e.g.~\citet{demarco2003}) that the discrepancy 
in the measurement of the cosmic ray flux by AGASA~\citep{agasa1998,agasa2000} 
and the monocular HiRes detector~\citep{prl2004} can be reconciled to some 
degree by an {\it ad hoc} change of about $30\,\%$ in the relative energy scales of the 
experiments.  In this scenario, numerically equal energy thresholds in HiRes and 
AGASA are not equivalent, and in order to compare the two data sets, one should
either raise the nominal AGASA threshold or lower the HiRes energy
threshold, or apply a combination of both shifts.  

We do not speculate on the nature of the flux discrepancy here.  At this 
point, there is no indication that it is caused by such a systematic 
shift in the energy scale of one or both of the experiments.  We also note that
lowering the energy threshold could potentially make magnetic bending a more 
serious limitation.
However, since the possibility of an energy shift has been suggested by 
others, it is useful to evaluate whether the conclusions of the maximum 
likelihood analysis change under this assumption.  Therefore, the method 
is applied again to the data set combining the 57 AGASA events above 
$4.0\times 10^{19}$\,eV and the 40 HiRes events above $3.0\times 10^{19}$\,eV, 
an energy threshold roughly $30\,\%$ lower.  

Lowering the HiRes energy cut admits an event that is near the aforementioned 
AGASA triplet.  This HiRes event has an energy of 37.6\,EeV and is located at 
right ascension $\alpha=169.0^{\circ}$ and declination $\delta=55.9^{\circ}$.  
The location with the largest $\ln\mathcal R$ is close to this event, 
at $\alpha=169.1^{\circ}\pm0.6^{\circ}$ and $\delta=56.3^{\circ}\pm0.4^{\circ}$.
The maximum likelihood ratio is $\ln\mathcal R=12.98$ for $n_{s}=3.9^{+2.2}_{-1.7}$.
47 out of $10^{4}$ random isotropic data sets have a larger value of maximum
$\ln\mathcal R$.

It is not possible to convert this number into a valid chance probability 
for the combination of the AGASA triplet and the additional HiRes event
because it contains several biases.  While the maximum 
likelihood method removes the bias due to choosing an angular bin size, it does not 
remove the bias due to choosing an energy threshold.  The AGASA threshold 
of $4.0\times 10^{19}$\,eV was chosen in~\citet{agasa1996} precisely 
because it maximized the clustering signal in the early data set, a data 
set which comprises the first 30 of the 57 AGASA events (and two of the 
triplet events).  An unbiased search for point sources using this energy 
threshold would require that these 30 events be set aside.  Further bias occurs 
because of the {\it ad hoc} lowering of the HiRes energy threshold.

The best estimate for the position can be considered an {\it a priori} 
location for a point source search with statistically independent future data.  
To illustrate the sensitivity of a new data set to this hypothesis, simulations
show that a simple binned analysis with a 
$1^{\circ}$ source bin radius centered on this location would give an unbiased
chance probability of $10^{-5}$ if 2 out of 40 future events
fall into the source bin.  In this calculation, it is assumed that the geometrical 
acceptance for the 40 new events is the same as for the original 40 events.

HiRes is currently the only detector observing the cosmic ray sky in this energy range
from the northern hemisphere.  We anticipate several more years of data taking, and 
the search for cosmic ray point sources at the highest energies will 
continue when more data becomes available.

\acknowledgments
The HiRes project is supported by the National Science Foundation under
contract numbers NSF-PHY-9321949, NSF-PHY-9322298, NSF-PHY-9974537,
NSF-PHY-0098826, NSF-PHY-0245428, by the Department of Energy Grant 
FG03-92ER40732, and by the Australian Research Council.  The cooperation 
of Colonels E. Fisher and G. Harter, the US Army and Dugway Proving Ground 
staff is appreciated.  The research of G. Farrar is supported in part by 
NSF-PHY-0101738 and NASA NAG5-9246; simulations were performed in part on
Mafalda, a linux cluster acquired with help from NSF-MRI-0116590.  We thank the 
authors of CORSIKA for providing us with the simulation code.

\end{document}